\documentclass[12pt,a4paper,reqno]{article}
\usepackage{amsmath}\usepackage{amssymb}
\usepackage{amsthm}

\renewcommand{\(}{\left(}
\renewcommand{\)}{\right)}

\newcommand{\e}{\text{e}}

\renewcommand{\c}{\mathbf{c}}

\newcommand{\A}{\mathbf{A}}

\newcommand{\1}{\mathbf{1}}

\newcommand{\x}{\times}
\newcommand{\<}{\langle}
\renewcommand{\>}{\rangle}
\newcommand{\half}{\tfrac{1}{2}}

\newcommand{\sqrthalf}{\frac{1}{\sqrt{2}}}
\newcommand{\third}{\tfrac{1}{3}}

\newcommand{\sixth}{\tfrac{1}{6}}
\newcommand{\quarter}{\tfrac{1}{4}}

\newcommand{\tr}{\operatorname{tr}}
\renewcommand{\Re}{\operatorname{Re}}

\theoremstyle{plain} %% This is the default 
\newtheorem{theorem}{Theorem}

\numberwithin{equation}{section}

\begin{document} 
\title{How entangled can two couples get?\footnote{See ref.\ \cite{CCGW}.}}
\author{\\\\A. Higuchi$^1$ and A. Sudbery$^2$\\
\small \emph{Dept. of Mathematics, University of York, 
Heslington, York, YO10 5DD, U.K.}
\\\small $^1$email: ah28@york.ac.uk 
\\\small $^2$email: as2@york.ac.uk}

\date{2 May 2000, revised 3 June 2000}

\maketitle

\begin{abstract} We describe a pure state of four qubits whose
single-qubit density matrices are all maximally mixed and whose average
entanglement as a system of two pairs of qubits
appears to be maximal.

\end{abstract}

\section{Introduction} \label{S:Intro}

In a system of two qubits the state
\begin{equation} 
\label{singlet} |C_2\> = \frac{1}{\sqrt{2}}\(|00\> + |11\>\)
\end{equation}
is, on all counts, the most entangled of all pure states. It gives the
greatest violation of Bell inequalities, it has the largest entropy of
entanglement, and its one-party reduced states are both maximally mixed.
All of these properties determine it uniquely up to local unitary
transformations.

A pure state of three qubits with similar properties is the GHZ state
\begin{equation} \label{GHZ} |C_3\> = \sqrthalf (|000\> + |111\>).
\end{equation}
This state has the maximum value of pure 3-party entanglement, as
measured by Wootters's 3-tangle \cite{Woot:quant, Woot:tang}, and its
one-particle reduced density matrices are all maximally mixed. Like the
two-qubit state $|C_2\>$, it is characterised uniquely, up to local
unitary transformations, by the latter property \cite{SchlienzMahler}.

The obvious $n$-party generalisation is the ``Schr\"odinger cat" state
\begin{equation} \label{cat_n} 
|C_n\> = \sqrthalf (|00\cdots 0\> + |11\cdots 1\>).
\end{equation}
Like $|C_2\>$ and $|C_3\>$, this state has the property that its
one-party reduced states are all maximally mixed. On the strength of
this, $|C_n\>$ is sometimes called the ``maximally entangled" pure state
of $n$ qubits. For $n>3$, however, not all states with this property are
locally equivalent, and it is not clear that $|C_n\>$ is really the most
entangled of them. Here we examine the case $n=4$, show that there are
4-qubit states which are more entangled than $|C_4\>$, and attempt to
find the most entangled among them.

A four-qubit system can be regarded, in three different ways, as a
system of two pairs of qubits, and one can ask how entangled are these
pairs. In the state $|C_4\>$ this entanglement is not maximal: each pair
$X,Y$ of the four qubits $A,B,C,D$ is not in the maximally mixed state 
but exhibits
correlations, all two-qubit density matrices being
\begin{equation} \label{rho2} \rho_{XY} = \half \(|00\>\<00| + |11\>\<11|\).
\end{equation}
But there do exist pure states of four qubits in which four of the six
two-qubit reduced density matrices are maximally mixed. (We note that
these density matrices come in pairs: if the reduced state of one pair
is maximally mixed, so is that of the complementary pair.) For example,
the state
\begin{equation} \label{example} 
|\Psi\> = \half\(|0000\> + |0111\> + |1001\> + |1110\>\)
\end{equation}
has two-qubit density matrices
\begin{align*}
\rho_{AB} &= \rho_{AC} = \rho_{BD} = \rho_{CD} = \quarter\1,\\
\rho_{AD} &= \half\(|++\>\<++| + |--\>\<--|\),\\
\rho_{BC} &= \half\(|00\>\<00| + |11\>\<11|\),
\end{align*}
where 
\begin{equation} \label{+-} |\pm\> = \frac{1}{\sqrt{2}}\(|0\> \pm |1\>\).
\end{equation}
The one-qubit reduced density matrices of $|\Psi\>$ are all maximally
mixed. But $|\Psi\>$ is not locally equivalent to the cat state
$|C_4\>$, for it has different values from that state of the 
entanglement entropies
$E_{AB}, E_{AC}, E_{BD}, E_{CD}$, which are invariants under local
unitary transformations. (We write
\begin{equation} \label{entropy} E_{XY} = - \tr(\rho_{XY}\log_2\rho_{XY}), 
\qquad \rho_{XY} = \tr_{ZW}|\Psi\>\<\Psi|
\end{equation}
where $\{W,X,Y,Z\}$ is a permutation of $\{A,B,C,D\}$). In fact
\cite{WuZhang}, the state $|\Psi\>$ cannot be asymptotically reversibly
converted into any collection of the states $|C_n\>$ for $n=2,3,4$.

It is natural to wonder whether there is a still more entangled state in
which the reduced density matrices of all six pairs of qubits are
maximally entangled.\footnote{This question has been studied by Gisin and
Bechmann-Pasquinucci \cite{GisinBP} for the general case of $n$ qubits,
but under the restriction that the states are symmetric under
permutations of the qubits. This is an unnatural requirement in this
context, since it is not invariant under local unitary transformations.}
In Section 2 we will show that there is no such state of four qubits.
The situation changes, however, if particles with larger state spaces
are considered, and we will show that a system of four four-state particles
does have pure states with this property.

In Section 3 we ask what is the maximum two-pair entanglement possible
for a system of four qubits. Taking as a measure the average
entanglement entropy $\<E_2\>=\third(E_{AB} + E_{AC} + E_{AD})$, we exhibit a
pure state $|M_4\>$ with a greater value of this quantity than the state
\eqref{example}; we show that $\<E_2\>$ has a stationary value at $|M_4\>$, 
and present
evidence that this is a global maximum.

\section{Non-existence of maximal entanglement\\ between all pairs of
pairs}

\begin{theorem} There is no pure state of four qubits whose two-qubit
density matrices are all multiples of the identity.
\end{theorem}

\begin{proof} Write the four-qubit state $|\Psi\>$ as
\[ 
  |\Psi\> = \sum_{i,j,k,l=0,1}t^{ijkl}|i\>|j\>|k\>|l\>
\]
The tensor $t^{ijkl}$ can be regarded as a $4\x4$ matrix in three
different ways:
\[
  t^{ijkl} = \half (U_1)^{ij}_{kl} = \half (U_2)^{ik}_{jl} 
  = \half (U_3)^{il}_{jk}
\]
and the requirement of maximal entanglement is that $U_1, U_2$ and $U_3$
should all be unitary matrices.

We show first that by local unitary transformations we can arrange that
the coordinates of $|\Psi\>$ satisfy
\[
  t^{1000} = t^{0100} = t^{0010} = t^{0001} = 0.
\]
To do this, we find normalised states
$|\alpha\>,|\beta\>,|\gamma\>,|\delta\>$ which maximise
$N=|\<\Psi|\alpha\>|\beta\>|\gamma\>|\delta\>|^2$. Such states certainly
exist, since $N(\alpha,\beta,\gamma,\delta)$ is a continuous function on
the compact space $S^3\x S^3\x S^3\x S^3$. Change basis in each of the
single-qubit spaces so that $|\alpha\>|\beta\>|\gamma\>|\delta\>$
becomes the basis state $|0000\>$; then $N=|t^{0000}|^2$. But if
$t^{1000}$ were non-zero we could change basis for states of the first
qubit so as to increase $|t^{0000}|$; since we have maximised it, we
must have $t^{1000}=0$. Similarly $t^{0100}=t^{0010}=t^{0001}=0$.

The matrices $U_1,U_2,U_3$ which have to be unitary are now
\footnotesize
\[
\begin{pmatrix}t^{0000}&0&0&t^{0011}\\0&t^{0101}&t^{0110}&t^{0111}\\
   0&t^{1001}&t^{1010}&t^{1011}\\t^{1100}&t^{1101}&t^{1110}&t^{1111}
\end{pmatrix},
\begin{pmatrix}t^{0000}&0&0&t^{0101}\\0&t^{0011}&t^{0110}&t^{0111}\\
   0&t^{1001}&t^{1100}&t^{1101}\\t^{1010}&t^{1011}&t^{1110}&t^{1111}
\end{pmatrix},
\begin{pmatrix}t^{0000}&0&0&t^{0110}\\0&t^{0011}&t^{0101}&t^{0111}\\
   0&t^{1010}&t^{1100}&t^{1110}\\t^{1001}&t^{1011}&t^{1101}&t^{1111}
\end{pmatrix}.
\]
\normalsize Hence the following three conditions must hold:

1. $\quad t^{0011} = 0\quad$ or $\quad t^{0111} = t^{1011} = 0$;

2. $\quad t^{0101} = 0\quad$ or $\quad t^{0111} = t^{1101} = 0$;

3. $\quad t^{0110} = 0\quad$ or $\quad t^{0111} = t^{1110} = 0$.

\noindent Following the branching consequences of these, and requiring
that the matrices are unitary, leads to a conclusion of the form that
three $2\x 2$ matrices
\[ \begin{pmatrix} a&b\\c&d \end{pmatrix},\qquad
   \begin{pmatrix} b&e\\f&c \end{pmatrix},\qquad
   \begin{pmatrix} a&e\\f&d \end{pmatrix}
\]
must all be unitary. In this situation all the matrix elements must be
non-zero: if, for example, $a=0$, then $d=0$ and $b,c,e,f$ would all
have unit modulus, which is impossible if the second matrix is to be
unitary. Now unitarity gives
\[
a = -\frac{b\overline{d}}{\overline{c}} = -\frac{e\overline{d}}{\overline{f}},
\]
so
\[
  b\overline{f} = e\overline{c}.
\]
But
\[
  b\overline{f} + e\overline{c} = 0,
\]
so $b\overline{f} = e\overline{c} = 0$, which is impossible. 
\end{proof}

Four-party states with this property do exist, however, 
if the individual parties
have more independent states. For example, suppose each party
has a 4-dimensional state space; then the (non-normalised) state with
coordinates
\[
  t^{ijkl} =  \begin{cases} 1 &\text{ if $i=j=k=l$ or $(i,j,k,l)$
is an even permutation of $(1,2,3,4)$}\\ 0 &\text{ otherwise }\end{cases}
\]
has every two-qubit density matrix equal to a multiple of the identity.

We note that the method introduced in this proof can be used to define a
canonical form for any multipartite pure state, in which a maximal number of
coordinates are zero. This can be regarded as a generalisation of the
Schmidt decomposition of a bipartite state. 
Details can be found in \cite{CHS}.

\section{A Maximally Entangled Four-qubit State?}

Given that a four-qubit state cannot have maximal entropy of entanglement
for every two-qubit subset, we now ask what is the greatest possible
average for such entropies, i.e. we seek to maximise
\begin{align*}
\<E_2\> &= \sixth (E_{AB} + E_{AC} + E_{AD} + E_{BC} +E_{BD} + E_{CD})\\
        &= \third (E_{AB} +E_{AC} + E_{AD}).
\end{align*}
The second equality holds because complementary pairs have equal entropy. 

We have not been able to solve this problem analytically. We will
adopt a heuristic approach, using an (indefensible) analogy with the
two-qubit system to obtain a candidate maximally entangled state and then
showing that $\<E_2\>$ is indeed stationary at this state and appears to
be maximal. 

Maximally entangled states of two qubits like \eqref{singlet} are
often referred to as ``singlet" states, since an example of such a state
is the state of two spin-$\half$ particles with zero total
angular momentum. This description is not invariant under local unitary
transformations, which need not preserve the total angular momentum.
Nevertheless, let us take this as a hint in investigating four-qubit
states. A system of four spin-$\half$ particles has two independent
singlet states. The most symmetric combination of these, in the sense
that all of its two-qubit density matrices are unitarily equivalent, is
locally equivalent to 
\[
  |M_4\> = \frac{1}{\sqrt{6}}\left[ |0011\> + |1100\>
     +\omega (|1010\> + |0101\>) + \omega^2 (|1001\> + |0110\>)\right],
\]
where $\omega = e^{2\pi i/3}$, 
or to $|\overline{M}_4\>$, the complex conjugate of $|M_4\>$ in the 
computation basis.
Note that this state is not symmetric
between the qubits but belongs to a two-dimensional representation of
the permutation group, the other state in the 
representation being  $|\overline{M}_4\>$.

The two-qubit density matrices of $|M_4\>$ are
\begin{equation} \label{M4}
  \rho_{AB} = \rho_{AC} = \rho_{AD} = 
  \sixth \left[|00\>\<00| + |11\>\<11| + |\Phi_+\>\<\Phi_+|\right] +
  \half|\Phi_-\>\<\Phi_-|
\end{equation}
where 
\[ 
  |\Phi_\pm\> = \frac{1}{\sqrt{2}}(|10\> \pm |01\>)
\]
(in angular momentum terms, the two-qubit reduced states are equal
mixtures of a singlet and a maximally mixed triplet). Hence the
entanglement entropies are 
\[
  E_{AB} = E_{AC} = E_{AD} = 1 + \half\log_2 3.
\]
Comparing with the cat state $|C_4\>$, for which 
\[
  E_{AB} = E_{AC} = E_{AD} = 1,
\]
and the state of \eqref{example}, for which 
\[
  E_{AB} = E_{AC} = 2, \qquad E_{AD} = 1,
\]
we see that $|M_4\>$ has a greater value of $\<E_2\>$ than either. 

We will now show that the function $\<E_2\>$ is stationary at $|M_4\>$.
To simplify the calculation, we consider the functions $E_{XY}$ defined
by \eqref{entropy} for all state vectors $|\Psi\>$, though these
functions coincide with the entropies only when $|\Psi\>$ is normalised.
Suppose $|\Psi\>$ changes by $|\delta\Psi\>$. Because the trace makes the
operators behave as if they commuted, the consequent change in $E_{XY}$
is given by 
\[
  \delta E_{XY} = - \tr\left[\rho_{XY} \log_2(\e \rho_{XY})\right].
\]
At $|\Psi\> = \sqrt{6}|M_4\>$ we find
\begin{align*}
\delta E_{AB} &= \Re\left[ -\<\Psi|\delta\Psi\>\log_2(3\e^2) +
\<\overline{\Psi}|\delta\Psi\>\log_2 3\right],\\
\delta E_{AC} &= \Re\left[ - \<\Psi|\delta\Psi\>\log_2(3\e^2) + 
\omega\<\overline{\Psi}|\delta\Psi\>\log_2 3\right],\\
\delta E_{AD} &= \Re \left[ - \<\Psi|\delta\Psi\>\log_2(3\e^2) + 
\omega^2\<\overline{\Psi}|\delta\Psi\>\log_2 3\right],
\end{align*}
where $|\overline{\Psi}\>$ is the complex conjugate of $|\Psi\>$ in the
computation basis.
Hence the gradient
of the average $\<E_2\>$ in the real 32-dimensional space of all state
vectors is in the direction of $|\Psi\>$:
\[
  \delta \<E_2\> =  - \log_2(3\e^2)\Re\, \<\Psi|\delta \Psi\>.
\]
Thus $\<E_2\>$ is stationary at $|\Psi\>$ for variations on the sphere
of vectors with the same norm as $|\Psi\>$. But each $E_{XY}$, as given
by \eqref{entropy}, is linearly related to the true entropy of
entanglement, so the average two-party entropy is stationary at
$|M_4\>$.

We have searched numerically for states which maximise $\<E_2\>$ starting
from several arbitrarily chosen states.  All the states we have obtained 
in this manner are locally equivalent to  
$|M_4\>$ or $|\overline{M}_4\>$.

\section{Robustness of the entanglement}

One of the criteria for maximal entanglement suggested in \cite{GisinBP}
is that the entanglement should be maximally fragile: a
measurement on any one of the qubits destroys the entanglement between
the remaining qubits. This is true of the cat state \eqref{cat_n} 
if the measurement
projects onto the computation basis $\{|0\>, |1\>\}$. Projection onto the
basis $\{|+\>, |-\>\}$ defined in \eqref{+-}, however, leaves the
remaining qubits in a cat state. 

The state constructed in the previous section behaves in the opposite
way to the cat state: measurement of one qubit leaves the other three
qubits in an entangled state, and the amount of entanglement afterwards
is independent of what measurement is performed. To be
precise, projection of one qubit onto a computation basis state 
leaves the remaining qubits in one of the entangled states 
\begin{align} &\frac{1}{\sqrt{3}}\(|011\>+\omega |101\>
+\omega^2|110\>\)\\
\text{or}\qquad &\frac{1}{\sqrt{3}}\(|100\>+\omega |010\>
+\omega^2|001\>\).
\end{align}
These states,
which are clearly locally equivalent, have
recently been identified as having maximal average two-qubit
entanglement in a certain sense \cite{DurrVidalCirac}. Projection onto
any other state of the first qubit leaves the other three qubits in
another state which is locally equivalent to the above.
This follows from the fact that the state $|M_4\>$ is an SU$(2)$
singlet, so that the unitary transformation from the computation basis
to any other basis of one qubit can be undone by performing 
the same transformation
on the other three qubits. 

Thus the entanglement of the state $|M_4\>$ is
\emph{robust}. Any carelessness by the holder of any one of the qubits,
resulting in an uncontrolled decoherence of that qubit, does
not completely destroy the entanglement of the remaining qubits, and
always leaves them with the same amount of entanglement.

\section*{Acknowledgements}
AS is grateful for the hospitality of the Physics Department at Imperial
College, London, where this work was started, and to Martin Plenio for a
helpful conversation. 

\newpage
\providecommand{\bysame}{\leavevmode\hbox to3em{\hrulefill}\thinspace}

\end{document}